\documentclass[letterpaper]{article}
\usepackage[T1]{fontenc}
\usepackage[latin1]{inputenc}

\makeatletter

\providecommand{\LyX}{L\kern-.1667em\lower.25em\hbox{Y}\kern-.125emX\@}

\usepackage[T1]{fontenc}
\usepackage[latin1]{inputenc}

\makeatletter

\usepackage[T1]{fontenc}
\usepackage[latin1]{inputenc}
\usepackage{setspace}

\makeatletter

\usepackage[T1]{fontenc}

\makeatother
\makeatother

\begin{document}

\title{Some conceptual issues involving probability in quantum mechanics\thanks{
It is a pleasure to dedicate this article to Arthur Fine. The subject of our
paper is close to one of Arthur's best known articles on the foundations of
physics \cite{Fine}. 
} }

\author{\textbf{J. Acacio de Barros}\thanks{
On leave from: Departamento de F\'{\i}sica -- ICE, Universidade Federal de
Juiz de Fora, Juiz de Fora, MG 36036-330, Brazil. E-mail: barros@ockham.stanford.edu. 
} \textbf{~and Patrick Suppes}\thanks{
E-mail: suppes@ockham.stanford.edu. 
}\\
 CSLI -- Ventura Hall \\
 Stanford University \\
 Stanford, CA 94305-4115 }

\date{\today{}}

\maketitle

\section{Introduction}

The issue of the completeness of quantum mechanics has been a subject of intense
research for almost a century. One of the most influential papers is undoubtedly
that of Eintein, Podolski and Rosen \cite{EPR}, where after analyzing entangled
two-particle states they concluded that quantum mechanics could not be considered
a complete theory. In 1964 John Bell showed that not only was quantum mechanics
incomplete but, if one wanted a complete description of reality that was local,
one would obtain correlations that are incompatible with the ones predicted
by quantum mechanics \cite{Bell}. This happens because some quantum mechanical
states do not allow for the existence of joint probability distributions of
all the possible outcomes of experiments. If a joint distribution exists, then
one could consistently create a local hidden variable that would factor this
distribution. The nonexistence of local hidden variables that would ``complete''
quantum mechanics, hence the nonexistence of joint probability distributions,
was verified experimentally in 1982 by Aspect, Dalibard and Roger \cite{Aspect},
when they showed, in a series of beautifully designed experiments, that an entangled
photon state of the form 
\begin{equation}
\label{superposition}
|\psi \rangle =\frac{1}{\sqrt{2}}(|+-\rangle -|-+\rangle ),
\end{equation}
 (where \( |+-\rangle \equiv |+\rangle _{A}\otimes |-\rangle _{B} \) represents,
for example, two photons \( A \) and \( B \) with helicity \( +1 \) and \( -1, \)
respectively) violates the Clauser-Horne-Shimony-Holt form of Bell's inequalities
\cite{Clauseretal}, as predicted by quantum mechanical computations. More recently,
Weihs et al. confirmed Aspect's experiment with a truly random selection of
the polarization angles, thus with a more strict nonlocality criteria satisfied
\cite{Weihs}. We note that the proof that the Clauser et al. form of Bell's
inequalities implies the existence of a joint probability distribution of the
observable random variables is the mains result in \cite{Fine}.

The nonexistence of joint probability distributions also comes into play in
the consistent-history interpretation of quantum mechanics. In this interpretation,
each sequence of properties for a given quantum mechanical system represents
a possible history for this system, and a set of such histories is called a
family of histories \cite{gellmann}. A family of \emph{consistent} histories
is one that has a joint probability distribution for all possible histories
in this family, with the joint probability distribution defined as any probability
measure on the space of all histories. One can easily show that quantum mechanics
implies the nonexistence of such probability functions for some families of
histories. Families of histories that do not have a joint probability distribution
are called inconsistent histories.

Another important example, also related to the nonexistence of a joint probability
distribution, is the famous Kochen-Specker theorem, that shows that a given
hidden-variable theory that is consistent with the quantum mechanical results
has to be contextual \cite{kochenspecker}, i.e., the hidden variable has to
depend on the values of the actual experimental settings, regardless of how
far apart the actual components of the experiment are located (throughout this
paper, we will use interchangeably the concepts of local and noncontextual hidden
variables; for a detailed discussion, see \cite{suppeshiddenvariable} and \cite{despagnat}).

More recently, a marriage between Bell's inequalities and the Kochen-Specker
theorem led to the Greenberger-Horne-Zeilinger (GHZ) theorem. The GHZ theorem
shows that if one assumes that one can consistently assign values to the outcomes
of a measurement before the measure is performed, a mathematical contradiction
arises \cite{GHZ} --- once again, having a complete data table would allow
us to compute the joint probability distribution, so we conclude that no joint
distribution exists that is consistent with quantum mechanical results. In this
paper, we propose the usage of nonmonotonic upper probabilities as a tool to
derive consistent joint upper probabilities for the contextual hidden variables.

\section{The GHZ Theorem}

In 1989 Greenberger, Horne and Zeilinger (GHZ) proved that if the quantum mechanical
predictions for entangled states are correct, then the assumption that there
exist noncontextual hidden variables that can accommodate those predictions
leads to contradictions \cite{GHZ}. Their proof of the incompatibility of noncontextual
hidden variables with quantum mechanics is now known as the GHZ theorem. This
theorem proposes a new test for quantum mechanics based on correlations between
more than two particles. What makes the GHZ theorem distinct from Bell's inequalities
is the fact that they use only perfect correlations. The argument for the GHZ
theorem, as stated by Mermin \cite{Mermin}, goes as follows. We start with
a three-particle entangled state 
\begin{equation}
|\psi \rangle =\frac{1}{\sqrt{2}}(|+\rangle _{1}|+\rangle _{2}|-\rangle _{3}+|-\rangle _{1}|-\rangle _{2}|+\rangle _{3}),
\end{equation}
 where we use a notation similar to that of equation (\ref{superposition}).
This state is an eigenstate of the following spin operators: 
\begin{eqnarray}
\hat{\mathbf{A}} & = & \hat{\sigma }_{1x}\hat{\sigma }_{2y}\hat{\sigma }_{3y},\, \, \, \, \hat{\mathbf{B}}=\hat{\sigma }_{1y}\hat{\sigma }_{2x}\hat{\sigma }_{3y},\\
\hat{\mathbf{C}} & = & \hat{\sigma }_{1y}\hat{\sigma }_{2y}\hat{\sigma }_{3x},\, \, \, \, \hat{\mathbf{D}}=\hat{\sigma }_{1x}\hat{\sigma }_{2x}\hat{\sigma }_{3x}.
\end{eqnarray}
 If we compute the expected values for the correlations above, we obtain at
once that \( E(\hat{\mathbf{A}})=E(\hat{\mathbf{B}})=E(\hat{\mathbf{C}})=1 \)
and \( E(\hat{\mathbf{D}})=-1. \) Let us now suppose that the value of the
spin for each particle is dictated by a hidden variable \( \lambda  \), and
let us call this value \( s_{ij}(\lambda ), \) where \( i=1...3 \) and \( j=x,y. \)
Then, we have that 
\begin{eqnarray}
E(\hat{\mathbf{A}}\hat{\mathbf{B}}\hat{\mathbf{C}}) & = & (s_{1x}s_{2y}s_{3y})(s_{1y}s_{2x}s_{3y})(s_{1y}s_{2y}s_{3x})\label{Contradiction1} \\
 & = & s_{1x}s_{2x}s_{3x}(s^{2}_{1y}s^{2}_{2y}s^{2}_{3y}).
\end{eqnarray}
 Since the \( s_{ij}(\lambda ) \) can only be \( 1 \) or \( -1, \) we obtain
\begin{equation}
\label{Contradiction2}
E(\hat{\mathbf{A}}\hat{\mathbf{B}}\hat{\mathbf{C}})=s_{1x}s_{2x}s_{3x}=E(\hat{\mathbf{D}}).
\end{equation}
 But (\ref{Contradiction1}) implies that \( E(\hat{\mathbf{A}}\hat{\mathbf{B}}\hat{\mathbf{C}})=1 \)
whereas (\ref{Contradiction2}) implies \( E(\hat{\mathbf{A}}\hat{\mathbf{B}}\hat{\mathbf{C}})=E(\hat{\mathbf{D}})=-1, \)
a clear contradiction. It is clear from the above derivation that one could
avoid contradictions if we allowed the value of \( \lambda  \) to depend on
the experimental setup, i.e., if we allowed \( \lambda  \) to be a contextual
hidden variable. In other words, what the GHZ theorem proves is that noncontextual
hidden variables cannot reproduce quantum mechanical predictions.

This striking characteristic of GHZ's predictions, however, has a major problem.
How can one verify experimentally predictions based on correlation-one statements,
since experimentally one cannot obtain events perfectly correlated? This problem
was also present on Bell's original paper, where he considered cases where the
correlations were one. To ``avoid Bell's experimentally unrealistic restrictions'',
Clauser, Horne, Shimony and Holt \cite{Clauseretal} derived a new set of inequalities
that would take into account imperfections in the measurement process. However,
Bell's inequalities are quite different from the GHZ case, where it is necessary
to have experimentally unrealistic perfect correlations. This can be seen from
the following theorem (a version for a 4 particle entangled system is found
in \cite{PatAcacioGary}).

\begin{description}
\item [Theorem~1]Let \( \mathbf{A}, \) \( \mathbf{B}, \) and \( \mathbf{C} \) be
three \( \pm 1 \) random variables and let \\
 (i) \( E(\mathbf{A})=E(\mathbf{B})=E(\mathbf{C})=1, \)\\
 (ii) \( E(\mathbf{ABC})=-1, \)\\
 then (i) and (ii) imply a contradiction. 
\end{description}
\noindent \emph{Proof:} By definition 
\begin{equation}
E(\mathbf{A})=P(a)-P(\overline{a}),
\end{equation}
 where we use a notation where \( a \) is \( \mathbf{A}=1 \), \( \overline{a} \)
is \( \mathbf{A}=-1 \), and so on. Since \( 0\leq P(a),\, P(\overline{a})\leq 1 \),
it follows at once from (i) that 
\begin{equation}
\label{(1)}
P(a)=1
\end{equation}
 and similarly 
\begin{equation}
\label{(2)}
P(b)=P(c)=1.
\end{equation}
 Using again the definition of expectation and the inequalities \( P(\overline{a}bc)\leq P(\overline{a})=0, \)
etc., we have 
\begin{equation}
\label{(3)}
\begin{array}{lll}
E(\mathbf{ABC}) & = & P(abc)+P(\overline{ab}c)+P(a\overline{bc})+P(\overline{a}b\overline{c})\\
 & = & P(abc)-[P(\overline{a}bc)+P(a\overline{b}c)+P(ab\overline{c})+P(\overline{a}\overline{b}\overline{c})]\\
 & = & 1,
\end{array}
\end{equation}
 from (\ref{(1)}) and (\ref{(2)}), since all but the first term on the right
is 0, and thus by conservation of probability \( P(ABC)=1 \). But (\ref{(3)})
contradicts (ii).

It is important to note that if we could measure all the random variables simultaneously,
we would have a joint distribution. The existence of a joint probability distribution
is a necessary and sufficient condition for the existence of a noncontextual
hidden variable \cite{SuppesZannoti2}. Hence, if the quantum mechanical GHZ
correlations are obtained, then no noncontextual hidden variable exists. However,
this abstract version of the GHZ theorem still involves probability-one statements.
On the other hand, the correlations present in the GHZ state are so strong that
even if we allow for experimental errors, the non-existence of a joint distribution
can still be verified, as we show in the following theorem \cite{AcacioPatPRL}.

\begin{description}
\item [Theorem~2]If \( \mathbf{A}, \) \( \mathbf{B}, \) and \( \mathbf{C} \) are
three \( \pm 1 \) random variables, a joint probability distribution exists
for the given expectations \( E(\mathbf{A}), \) \( E(\mathbf{B}), \) \( E(\mathbf{C}), \)
and \( E(\mathbf{ABC}) \) if and only if the following inequalities are satisfied:
\begin{equation}
\label{inequality1}
-2\leq E(\mathbf{A})+E(\mathbf{B})+E(\mathbf{C})-E(\mathbf{ABC})\leq 2,
\end{equation}
\begin{equation}
-2\leq -E(\mathbf{A})+E(\mathbf{B})+E(\mathbf{C})+E(\mathbf{ABC})\leq 2,
\end{equation}
\begin{equation}
-2\leq E(\mathbf{A})-E(\mathbf{B})+E(\mathbf{C})+E(\mathbf{ABC})\leq 2,
\end{equation}
\begin{equation}
\label{inequality4}
-2\leq E(\mathbf{A})+E(\mathbf{B})-E(\mathbf{C})+E(\mathbf{ABC})\leq 2.
\end{equation}

\end{description}
\emph{Proof:} First we prove necessity. Let us assume that there is a joint
probability distribution consisting of the eight atoms \( abc, \) \( ab\overline{c}, \)
\( a\overline{b}c, \) \( ...\overline{a}\overline{b}\overline{c} \). Then,
\[
E(\mathbf{A})=P(a)-P(\overline{a}),\]
 where 
\[
P(a)=P(abc)+P(a\overline{b}c)+P(ab\overline{c})+P(a\overline{b}\overline{c}),\]
 and 
\[
P(\overline{a})=P(\overline{a}bc)+P(\overline{a}\overline{b}c)+P(\overline{a}b\overline{c})+P(\overline{a}\overline{b}\overline{c}).\]
 Similar equations hold for \( E(\mathbf{B}) \) and \( E(\mathbf{C}). \) For
\( E(\mathbf{ABC}) \) we obtain 
\begin{eqnarray*}
E(\mathbf{ABC}) & = & P(\mathbf{ABC}=1)-P(\mathbf{ABC}=-1)\\
 & = & P(abc)+P(a\overline{b}\overline{c})++P(\overline{a}\overline{b}c)+P(\overline{a}b\overline{c})\\
 &  & -[P(a\overline{b}c)+P(ab\overline{c})+P(\overline{a}bc)+P(\overline{a}\overline{b}\overline{c})].
\end{eqnarray*}
 Corresponding to the first inequality above, we now sum over the probability
expressions for the expectations 
\[
F=E(\mathbf{A})+E(\mathbf{B})+E(\mathbf{C})-E(\mathbf{ABC}),\]
 and obtain the expression 
\begin{eqnarray}
F & = & 2[P(abc)+P(\overline{a}bc)+P(a\overline{b}c)+P(ab\overline{c})]\nonumber \\
 &  & -2[P(\overline{a}\overline{b}\overline{c})+P(\overline{a}\overline{b}c)+P(\overline{a}b\overline{c})+P(a\overline{b}\overline{c})],\nonumber 
\end{eqnarray}
 and since all the probabilities are nonnegative and sum to \( \leq 1 \), we
infer at once inequality (\ref{inequality1}). The derivation of the other three
inequalities is very similar.

To prove the converse, i.e., that these inequalities imply the existence of
a joint probability distribution, is slightly more complicated. We restrict
ourselves to the symmetric case 
\[
P(a)=P(b)=P(c)=p,\]

\[
P(\mathbf{ABC}=1)=q\]
 and thus 
\[
E(\mathbf{A})=E(\mathbf{B})=E(\mathbf{C})=2p-1,\]

\[
E(\mathbf{ABC})=2q-1.\]
 In this case, (\ref{inequality1}) can be written as 
\[
0\leq 3p-q\leq 2,\]
 while the other three inequalities yield just \( 0\leq p+q\leq 2 \). Let 
\[
x=P(\overline{a}bc)=P(a\overline{b}c)=P(ab\overline{c}),\]

\[
y=P(\overline{a}\overline{b}c)=P(\overline{a}b\overline{c})=P(a\overline{b}\overline{c}),\]

\[
z=P(abc),\]
 and 
\[
w=P(\overline{a}\overline{b}\overline{c}).\]
 It is easy to show that on the boundary \( 3p=q \) defined by the inequalities
the values \( x=0, \) \( y=q/3, \) \( z=0, \) \( w=1-q \) define a possible
joint probability distribution, since \( 3x+3y+z+w=1 \). On the other boundary,
\( 3p=q+2 \) a possible joint distribution is \( x=(1-q)/3, \) \( y=0, \)
\( z=q, \) \( w=0 \). Then, for any values of \( q \) and \( p \) within
the boundaries of the inequality we can take a linear combination of these distributions
with weights \( (3p-q)/2 \) and \( 1-(3p-q)/2 \), chosen such that the weighed
probabilities add to one, and obtain the joint probability distribution: 
\begin{eqnarray*}
x & = & \left( 1-\frac{3p-q}{2}\right) \frac{1-q}{3},\\
y & = & \left( \frac{3p-q}{2}\right) \frac{q}{3},\\
z & = & \left( 1-\frac{3p-q}{2}\right) q,\\
w & = & \frac{3p-q}{2}\left( 1-q\right) ,
\end{eqnarray*}
 which proves that if the inequalities are satisfied a joint probability distribution
exists, and therefore a noncontextual hidden variable as well, thus completing
the proof. The generalization to the asymmetric case is tedious but straightforward.

As a consequence of the inequalities above, one can show that the correlations
present in the GHZ state are so strong that even if we allow for experimental
errors, the non-existence of a joint distribution can still be verified \cite{AcacioPatPRL}.

\begin{description}
\item [Corollary]Let \( \mathbf{A}, \) \( \mathbf{B}, \) and \( \mathbf{C} \) be
three \( \pm 1 \) random variables such that\\
 (i) \( E(\mathbf{A})=E(\mathbf{B})=E(\mathbf{C})\geq 1-\epsilon  \), \\
 (ii) \( E(\mathbf{ABC})\leq -1+\epsilon  \), \\
 where \( \epsilon  \) represents a decrease of the observed \( GHZ \) correlations
due to experimental errors. Then, there cannot exist a joint probability distribution
of \( \mathbf{A}, \) \( \mathbf{B}, \) and \( \mathbf{C} \) if 
\begin{equation}
\epsilon <\frac{1}{2}.
\end{equation}

\end{description}
\emph{Proof:} To see this, let us compute the value of \( F \) define above.
We obtain at once that 
\[
F=3(1-\epsilon )-(-1+\epsilon ).\]
 But the observed correlations are only compatible with a noncontextual hidden
variable theory if \( F\leq 2 \), hence \( \epsilon <\frac{1}{2}. \) Then,
there cannot exist a joint probability distribution of \( \mathbf{A}, \) \( \mathbf{B}, \)
and \( \mathbf{C} \) satisfying (i) and (ii) if 
\begin{equation}
\epsilon <\frac{1}{2}.
\end{equation}

From the inequality obtained above, it is clear that any experiment that obtains
GHZ-type correlations stronger than \( 0.5 \) cannot have a joint probability
distribution. For example, the recent experiment made at Innsbruck \cite{Innsbruck}
with three-photon entangled states supports the quantum mechanical result that
no noncontextual hidden variable exists that explain their correlations \cite{AcacioPatPRL}.
Thus, with this reformulation of the GHZ theorem it is possible to use strong,
yet imperfect, experimental correlations to prove that a noncontextual hidden-variable
theory is incompatible with the experimental results.

\section{Upper and Lower Probabilities and the GHZ theorem}

We saw at the previous section that quantum mechanics does not allow, for some
cases, the definition of a joint probability distribution for all the observables.
However, if we weaken the probability axioms, it is possible to prove that one
can find a consistent set of upper probabilities for the events \cite{SuppesZannoti1}.
Upper probabilities are defined in the following way. Let \( \Omega  \) be
a nonempty set, \( F \) a boolean algebra on \( \Omega  \) and \( P^{*} \)
a real valued function on \( F. \) Then the triple \( (\Omega ,F,P^{*}) \)
is an \emph{upper probability} if for all \( \xi _{1} \) and \( \xi _{2} \)
in \( F \) we have that

\begin{description}
\item [(i)]\( 0\leq P^{*}(\xi _{1})\leq 1, \)
\item [(ii)]\( P^{*}(\emptyset )=0, \)
\item [(iii)]\( P^{*}(\Omega )=1, \)
\end{description}
and if \( \xi _{1} \) and \( \xi _{2} \) are disjoint, i.e. \( \xi _{1}\cap \xi _{2}=\emptyset , \)
then

\begin{description}
\item [(iv)]\( P^{*}(\xi _{1}\cup \xi _{2})\leq P^{*}(\xi _{1})+P^{*}(\xi _{2}). \)
\end{description}
As we can see, this last property weakens the standard axioms for probability,
as one of the consequences of these axioms is that it may be true, for an upper
probability, that

\[
\xi _{1}\subseteq \xi _{2}\mbox {\, and\, }P^{*}(\xi _{1})>P^{*}(\xi _{2}),\]
 a quite nonstandard property. In a similar way, \emph{lower probabilities}
are defined as satisfying the triple \( (\Omega ,F,P_{*}) \) such that for
all \( \xi _{1} \) and \( \xi _{2} \) in \( F \) we have that

\begin{description}
\item [(i)]\( 0\leq P_{*}(\xi _{1})\leq 1, \)
\item [(ii)]\( P_{*}(\emptyset )=0, \)
\item [(iii)]\( P_{*}(\Omega )=1, \)
\end{description}
and if \( \xi _{1} \) and \( \xi _{2} \) are disjoint, i.e. \( \xi _{1}\cap \xi _{2}=\emptyset , \)
then

\begin{description}
\item [(iv)]\( P_{*}(\xi _{1}\cup \xi _{2})\geq P_{*}(\xi _{1})+P_{*}(\xi _{2}). \)
\end{description}
Let us see how upper and lower probabilities can be used to obtain joint upper
and lower probability distributions. We can start with the standard Bell's variables
\( \mathbf{X}, \) \( \mathbf{Y} \) and \( \mathbf{Z}, \) where each random
variable represents a different angles for the Stern-Gerlach apparatus (we follow
the example in \cite{SuppesZannoti1}). In the experimental setup used by Bell,
a two-particle system with entangled spin state was used, and for that reason
we can only measure two variables at the same time. However, since they are
spin measurements, we have the constraint 
\[
P(\mathbf{X}=1)=P(\mathbf{Y}=1)=P(\mathbf{Z}=1)=\frac{1}{2}.\]
 The question that Bell posed is whether we can fill the missing values of the
data table in a way that is consistent with the correlations given by quantum
mechanics for the pairs of variables, that is, \( E(\mathbf{XY}), \) \( E(\mathbf{XZ}), \)
\( E(\mathbf{YZ}). \) It is well known that for some sets of angles, the joint
probability distribution of \( \mathbf{X}, \) \( \mathbf{Y}, \) and \( \mathbf{Z} \)
exists, while for other set of angles it does not exist. We can prove that the
joint doesn't exist in the following way. We start with the values for the correlations
used by Bell: 
\begin{eqnarray}
E(\mathbf{XY}) & = & -\frac{\sqrt{3}}{2},\label{marginals} \\
E(\mathbf{XZ}) & = & -\frac{\sqrt{3}}{2},\\
E(\mathbf{YZ}) & = & -\frac{1}{2}.
\end{eqnarray}
 The correlations above correspond to the angles \( \widehat{\mathbf{XY}}=30^{\mbox {o}}, \)
\( \widehat{\mathbf{YZ}}=30^{\mbox {o}} \) and \( \widehat{\mathbf{XZ}}=60^{\mbox {o}} \)
for the detectors, and require that 
\begin{eqnarray}
E(\mathbf{XY}) & = & E(\mathbf{XY}|\mathbf{Z}=1)P(\mathbf{Z}=1)+E(\mathbf{XY}|\mathbf{Z}=-1)P(\mathbf{Z}=-1),\nonumber \\
E(\mathbf{XZ}) & = & E(\mathbf{XZ}|\mathbf{Y}=1)P(\mathbf{Y}=1)+E(\mathbf{XZ}|\mathbf{Y}=-1)P(\mathbf{Y}=-1),\nonumber \\
E(\mathbf{YZ}) & = & E(\mathbf{YZ}|\mathbf{X}=1)P(\mathbf{X}=1)+E(\mathbf{YZ}|\mathbf{X}=-1)P(\mathbf{X}=-1),\nonumber 
\end{eqnarray}
 which can be written as 
\begin{eqnarray}
2E(\mathbf{XY}) & = & E(\mathbf{XY}|\mathbf{Z}=1)+E(\mathbf{XY}|\mathbf{Z}=-1),\label{system} \\
2E(\mathbf{XZ}) & = & E(\mathbf{XZ}|\mathbf{Y}=1)+E(\mathbf{XZ}|\mathbf{Y}=-1),\\
2E(\mathbf{YZ}) & = & E(\mathbf{YZ}|\mathbf{X}=1)+E(\mathbf{YZ}|\mathbf{X}=-1),
\end{eqnarray}
 because \( P(\mathbf{Z}=1)=P(\mathbf{Z}=-1), \) etc. Symmetry requires that
\begin{eqnarray}
E(\mathbf{XY}|\mathbf{Z}=1) & = & E(\mathbf{YZ}|\mathbf{X}=1),\\
E(\mathbf{XY}|\mathbf{Z}=-1) & = & E(\mathbf{YZ}|\mathbf{X}=-1)\label{systemfinal} 
\end{eqnarray}
 and if we use the requirement that all probabilities must sum to one we have
six equations and six unknown conditional expectations. It is easy to see that
the system of linear equations (\ref{system})---(\ref{systemfinal}) does not
have a solution for the correlations shown in (\ref{marginals}), hence no joint
probability distribution exists. What happened? The correlations are too strong
for us to fill up a table with all the experimental results, including the ones
that did not occur. One extreme example can be obtained if we use the extreme
case of correlation one expectations, given by 
\begin{eqnarray*}
E(\mathbf{XY}) & = & -1,\\
E(\mathbf{YZ}) & = & -1,\\
E(\mathbf{XZ}) & = & -1,
\end{eqnarray*}
 where once again no joint probability distribution exists.

What changes with upper probabilities? The system of linear equations (\ref{system})
becomes a system of inequalities: 
\begin{eqnarray}
2E^{*}(\mathbf{XY}) & \geq  & E^{*}(\mathbf{XY}|\mathbf{Z}=1)+E^{*}(\mathbf{XY}|\mathbf{Z}=-1),\label{systemupper} \\
2E^{*}(\mathbf{XZ}) & \geq  & E^{*}(\mathbf{XZ}|\mathbf{Y}=1)+E^{*}(\mathbf{XZ}|\mathbf{Y}=-1),\\
2E^{*}(\mathbf{YZ}) & \geq  & E^{*}(\mathbf{YZ}|\mathbf{X}=1)+E^{*}(\mathbf{YZ}|\mathbf{X}=-1),
\end{eqnarray}
 plus the symmetry 
\begin{eqnarray}
E^{*}(\mathbf{XY}|\mathbf{Z}=1) & = & E^{*}(\mathbf{YZ}|\mathbf{X}=1),\\
E^{*}(\mathbf{XY}|\mathbf{Z}=-1) & = & E^{*}(\mathbf{YZ}|\mathbf{X}=-1),\label{systemupperfinal} 
\end{eqnarray}
 and the fact that the sum of all upper probabilities must be greater or equal
than one. It is straightforward to obtain solutions to (\ref{systemupper})--(\ref{systemupperfinal}),
and then we can find upper probabilities that are consistent with the conditional
expectations.

The following theorem shows that the GHZ theorem fail if we allow lower probabilities.

\begin{description}
\item [Theorem~3]Let \( \mathbf{A}, \) \( \mathbf{B}, \) and \( \mathbf{C} \) be
three \( \pm 1 \) random variables and let \\
 (i) \( E_{*}(\mathbf{A})=E(\mathbf{A})=1, \)\\
 (ii) \( E_{*}(\mathbf{B})=E(\mathbf{B})=1, \)\\
 (iii) \( E_{*}(\mathbf{C})=E(\mathbf{C})=1, \)\\
 (iv) \( E_{*}(\mathbf{ABC})=E(\mathbf{ABC})=-1. \)\\
 Then, there exist a lower joint probability distribution that is compatible
with (i)---(iv). 
\end{description}
\emph{Proof:} We will prove this theorem by explicitly constructing a lower
joint probability distribution. First, we note that 
\[
E_{*}(\mathbf{A})=P_{*}(a)-P_{*}(\overline{a})=1,\]

\[
E_{*}(\mathbf{B})=P_{*}(b)-P_{*}(\overline{b})=1,\]

\[
E_{*}(\mathbf{C})=P_{*}(c)-P_{*}(\overline{c})=1,\]
 and hence

\begin{eqnarray}
P_{*}(a)=1, &  & P_{*}(\overline{a})=0,\label{uppers1} \\
P_{*}(b)=1, &  & P_{*}(\overline{b})=0,\\
P_{*}(c)=1 &  & P_{*}(\overline{c})=0.\label{uppers3} 
\end{eqnarray}
 From the definition of lowers and from (\ref{uppers1})--(\ref{uppers3}) we
have 
\begin{eqnarray}
P_{*}(abc)+P_{*}(a\overline{b}c)+P_{*}(ab\overline{c})+P_{*}(a\overline{b}\overline{c}) & \leq  & 1,\label{ineq1} \\
P_{*}(abc)+P_{*}(\overline{a}bc)+P_{*}(ab\overline{c})+P_{*}(\overline{a}b\overline{c}) & \leq  & 1,\\
P_{*}(abc)+P_{*}(\overline{a}bc)+P_{*}(a\overline{b}c)+P_{*}(\overline{a}\overline{b}c) & \leq  & 1,\label{ineq2} 
\end{eqnarray}
 and from (iv) 
\begin{eqnarray}
P_{*}(abc)+P_{*}(\overline{a}\overline{b}c)+P_{*}(a\overline{b}\overline{c})+P_{*}(\overline{a}b\overline{c})+ &  & \\
-P_{*}(\overline{a}bc)-P_{*}(a\overline{b}c)-P_{*}(ab\overline{c})-P_{*}(\overline{a}\overline{b}\overline{c}) & = & -1.\label{uppercorrelation} 
\end{eqnarray}
 The lowers must also be superadditive in the whole probability space, and we
have 
\begin{eqnarray}
P_{*}(abc)+P_{*}(\overline{a}\overline{b}c)+P_{*}(a\overline{b}\overline{c})+P_{*}(\overline{a}b\overline{c})+ &  & \\
P_{*}(\overline{a}bc)+P_{*}(a\overline{b}c)+P_{*}(ab\overline{c})+P_{*}(\overline{a}\overline{b}\overline{c}) & \leq  & 1.\label{totalupper} 
\end{eqnarray}
 From (\ref{uppercorrelation}) and (\ref{totalupper}) we have 
\[
P_{*}(abc)=P_{*}(\overline{a}\overline{b}c)=P_{*}(a\overline{b}\overline{c})=P_{*}(\overline{a}b\overline{c})=0\]
 and the system reduces to 
\begin{eqnarray}
P_{*}(a\overline{b}c)+P_{*}(ab\overline{c}) & \leq  & 1,\label{finalset1} \\
P_{*}(\overline{a}bc)+P_{*}(ab\overline{c}) & \leq  & 1,\\
P_{*}(\overline{a}bc)+P_{*}(a\overline{b}c) & \leq  & 1,\label{b} \\
P_{*}(\overline{a}bc)+P_{*}(a\overline{b}c)+P_{*}(ab\overline{c})+P_{*}(\overline{a}\overline{b}\overline{c}) & = & 1.\label{finalset2} 
\end{eqnarray}
 A possible solution for the system (\ref{finalset1})--(\ref{finalset2}) is
\begin{eqnarray*}
P_{*}(\overline{a}bc)=P_{*}(a\overline{b}c)=P_{*}(ab\overline{c}) & = & \frac{1}{3}\\
P_{*}(\overline{a}\overline{b}\overline{c}) & = & 0,
\end{eqnarray*}
 as we wanted to prove. In a similar way, we have the following:

\begin{description}
\item [Theorem~4]Let \( \mathbf{A}, \) \( \mathbf{B}, \) and \( \mathbf{C} \) be
three \( \pm 1 \) random variables and let \\
 (i) \( E^{*}(\mathbf{A})=E(\mathbf{A})=1, \)\\
 (ii) \( E^{*}(\mathbf{B})=E(\mathbf{B})=1, \)\\
 (iii) \( E^{*}(\mathbf{C})=E(\mathbf{C})=1, \)\\
 (iv) \( E^{*}(\mathbf{ABC})=E(\mathbf{ABC})=-1. \)\\
 Then, there exist an upper probability distribution that is compatible with
(i)---(iv). 
\end{description}
\emph{Proof:} Similar to the proof for the lower.

We note that the nonmonotonic upper and lower probabilities shown to exist in
Theorems 3 and 4 do not, because of their nonmonotonicity, satisfy the usual
definitional relation between upper and lower probabilities, for any event \( A \):
\[
P^{*}(A)=1-P_{*}(\overline{A}).\]

\section{Final Remarks}

To apply the upper probabilities to the GHZ theorem, we gave a probabilistic
random variable version of it. We then showed that, if we use upper probabilities,
the GHZ theorem does not hold anymore, and hence the inconsistencies cannot
be proved to exist for the upper probabilities. Such upper probabilities are
a natural way to deal with contextual problems in statistics. Whether they lead
to fruitful theoretical developments in a new direction is, however, an open
question.


\begin{thebibliography}{Gell-Mann and Hartle 1990}
\bibitem[Aspect at al. 1982]{Aspect}A. Aspect, J. Dalibard, and G. Roger, ``Experimental test of Bell's inequalities
using time-varying analyzers'', \emph{Phys. Rev. Lett.} \textbf{49}, 1804 (1982). 
\bibitem[Barros and Suppes 2000]{AcacioPatPRL}J. Acacio de Barros and Patrick Suppes, ``Inequalities for dealing with detector
inefficiencies in GHZ-type experiments'', to appear in \emph{Phys. Rev. Lett.,}
Jan. 24, 2000. 
\bibitem[Bell 1987]{Bell}J. S. Bell, ``On the Einstein-Podolski-Rosen paradox'', \emph{Physics} \textbf{1},
195 (1964). 
\bibitem[Bouwemeester et al. 1999]{Innsbruck}D. Bouwmeester, J-W. Pan, M. Daniell, H. Weingurter, and A. Zeilinger, ``Observation
of three-photon Greenberger-Horne-Zeilinger entanglement'', \emph{Phys. Rev.
Lett.} \textbf{82}, 1345 (1999). 
\bibitem[Clauser et al. 1969]{Clauseretal}J. F. Clauser, M. A. Horne, A. Shimony, and R. A. Holt, ``Proposed experiment
to test local hidden variable theories'', \emph{Phys. Rev. Lett.} \textbf{23},
880 (1969). 
\bibitem[D'Espagnat 1989]{despagnat}B. D'Espagnat, ``Nonseparability and the tentative descriptions of reality'',
in \emph{Quantum Theory and Pictures of Reality}, edited by W. Schommers (Springer-Verlag,
Berlin, 1989).
\bibitem[Einstein et al. 1935]{EPR}A. Einstein, B. Podolski, N. Rosen, ``Can the quantum mechanical description
of physical reality be considered complete?'', \emph{Phys. Rev.} \textbf{47},
777 (1935). 
\bibitem[Fine 1982]{Fine}A. Fine, ``Hidden variables, joint probability, and the Bell inequalities'',
\emph{Phys. Rev. Lett.} \textbf{48}, 291 (1982). 
\bibitem[Gell-Mann and Hartle 1990]{gellmann}M. Gell-Mann and J. B. Hartle, ``Quantum mechanics in the light of quantum
cosmology'', in \emph{Proceedings of the 3rd intenational symposium on Foundations
of Quantum Mechanics}, edited by S. Kobayashi, H. Ezawa, Y. Murayama and S.
Nomura (Tokyo, Japan Phys. Soc., Japan 1990). 
\bibitem[Greenberger et al. 1989]{GHZ}D. M. Greenberger, M. Horne, and A. Zeillinger, ``Going beyond Bell's theorem''
in \emph{Bell's Theorem, Quantum Theory, and Conceptions of the Universe}, edited
by M. Kafatos (Kluwer, Dordrecht, 1989). 
\bibitem[Kochen and Specker 1967]{kochenspecker}S. Kochen and E. P. Specker, ``The problem of hidden variables in quantum mechanics'',
\emph{Journal of Mathmatics and Mechanics} \textbf{17}, 59 (1967). 
\bibitem[Mermin 1990a]{Mermin}N. D. Mermin, ``Quantum misteries revisited'', \emph{Am. J. Phys.} \textbf{58},
731 (1990). 
\bibitem[Mermin 1990b]{Mermin2}N. D. Mermin, ``Extreme quantum entanglement in a superposition of macroscopically
distinct states'', \emph{Phys. Rev. Lett.} \textbf{65}, 1838 (1990). 
\bibitem[Peres 1995]{Peres}A. Peres, \emph{Quantum Theory: Concepts and Methods} (Kluwer Academic Pub.,
Dordrecht, 1995). 
\bibitem[Suppes et al. 1998]{PatAcacioGary}P. Suppes, J. Acacio de Barros, and G. Oas, ``A collection of probabilistic
hidden-variable theorems and counterexamples'' in \emph{Waves, Information
and Foundations of Physics}, edited by Riccardo Pratesi and Laura Ronchi (Italian
Physics Society, Bologna, 1998). 
\bibitem[Suppes and Zanotti 1976]{suppeshiddenvariable}P. Suppes and M. Zanotti, ``On the determinism of hidden variable theories
with strict correlation and conditional statistical independence of observables'',
in \emph{Logic and Probability in Quantum Mechanics}, edited by P. Suppes (Reidel,
Dordrecht, 1976).
\bibitem[Suppes and Zanotti 1981]{SuppesZannoti2}P. Suppes and M. Zanotti, ``When are probabilistic explanations possible?'',
\emph{Synthese} \textbf{48}, 191 (1981). 
\bibitem[Suppes and Zanotti 1991]{SuppesZannoti1}P. Suppes and M. Zanotti, ``Existence of hidden variables having only upper
probabilities'', \emph{Found. Phys}. \textbf{21}, 1479 (1991). 
\bibitem[Weihs et al. 1998]{Weihs}G. Weihs, T. Jennewein, C. Simon, H. Weinfurter, and A. Zeilinger, ``Violation
of Bell's inequality under strict Einstein locality conditions'', \emph{Phys.
Rev. Lett.} \textbf{81,} 5039 (1998). 
\end{thebibliography}
\end{document}